\input psfig
\documentstyle[twoside,fleqn,espcrc2]{article}

\newcommand{\AmS}{{\protect\the\textfont2
  A\kern-.1667em\lower.5ex\hbox{M}\kern-.125emS}}

\title{Quenched chiral logs, the $\eta'$ mass,
       and the hairpin diagram}

\author{A. Duncan\address{Dept. of Physics and Astronomy,University of Pittsburgh, Pittsburgh, PA 15260},%
        E,~Eichten\address{Fermilab, P.O. Box 500, Batavia, IL 60510}%
        S.~Perrucci$^{\rm c}$
        and 
        H.~Thacker\address{Dept. of Physics, University of Virginia, Charlottesville, VA 22901}}

\begin{document}

\begin{abstract}
Limits on the size of quenched chiral logs in the pion mass for Wilson fermions are investigated. 
The smallness of chiral logs is shown to be a result of the suppression of the hairpin diagram
for small $p^2$, such that the value of the hairpin on the pion mass shell is much smaller than
the physical $m_{\eta'}^2$. A direct calculation of the topological susceptibility from the same
data gives $m_{\eta'}\approx 1$ GeV.
\end{abstract}

\maketitle

\section{INTRODUCTION}

One of the most direct manifestations of closed quark loop effects
is in the mass of the flavor singlet
pseudoscalar $\eta'$ meson. The $\eta'$ propagator includes not only the valence
quark-antiquark term that appears in the nonsinglet propagator, but also the
contribution from diagrams in which the valence quark and antiquark annihilate.
These ``hairpin'' diagrams probe the topological structure of the gauge field via the
axial $U(1)$ anomaly. 
In a large $N_c$ approximation, the axial anomaly may be introduced perturbatively
as a term which breaks the $U(1)_A$ symmetry of the massless quark Lagrangian. From the
chiral Lagrangian point of view, the anomaly adds a term ${\cal L}_1$ to the 
$U(N_f)\times U(N_f)$ Lagrangian ${\cal L}_0$
which gives the $\eta'$ its mass. For low momentum,
and keeping only terms quadratic in the $\eta'$ field,
the most general form of such a term is (c.f. Ref.\cite{Sharpe})
\begin{equation}
\label{eq:L1}
{\cal L}_1 = \frac{1}{2}\left((A-1)\left[(\partial_{\mu}\eta')^2-m_{\pi}^2\eta'^2)\right] - Am_0^2 \eta'^2\right)
\end{equation}
This combines with the term in ${\cal L}_0$ to give an $\eta'$ mass of $m_{\eta'}=\sqrt{m_0^2+m_{\pi}^2}$.
The constant $A$ is a renormalization of the $\eta'$ field induced by the 
inclusion of the anomaly.
In the large $N_c$ framework, the term (\ref{eq:L1}) may be identified with the quenched
hairpin diagram. It corresponds to an amputated hairpin vertex of the form
\begin{equation}
\label{eq:hairvertex}
\Pi(p^2) = -(A-1)(p^2-m_{\pi}^2) + Am_0^2
\end{equation}
The main difference between quenched and full QCD is that in 
quenched QCD, the hairpin vertex appears only once in the $\eta'$ propagator,
while in full QCD it appears an arbitrary number of times. In the latter case,
the hairpin insertions sum up geometrically and shift the 
$p^2=m_{\pi}^2$ Goldstone pole to a pole at $p^2 = m_{\eta'}^2$. 
By contrast, the {\it quenched} $\eta'$ propagator 
includes only a single hairpin insertion. Not only is the Goldstone pole not 
cancelled, but the hairpin graph adds a double pole $1/(p^2-m_{\pi}^2)^2$ term to the 
propagator. Note that, by (\ref{eq:hairvertex}), the quenched diagram with a single hairpin insertion includes both a single pole and a double pole term, with coefficients
$1-A$ and $Am_0^2$ respectively. 

The appearance of a double pole in the quenched $\eta'$ propagator gives rise to
anomalous chiral behavior, e.g. in the relation
between the pion mass and the quark mass\cite{Sharpe,BG}.
The chiral symmetry result that $m_{\pi}^2$ is linear in the quark
mass is replaced in the quenched approximation by
\begin{equation}
m_{\pi}^2 \propto m_q^{\frac{1}{1+\delta}}
\end{equation}
where the parameter $\delta$ which determines the anomalous power behavior
is the coefficient of the quenched chiral log in the one-loop graph, and is
proportional to $Am_0^2$, the value of the hairpin insertion at $p^2
\approx m_{\pi}^2$. This gives
\begin{equation}
\delta = \frac{Am_0^2}{24\pi^2f_{\pi}^2}
\end{equation}
If we assume that $A\approx 1$, a rough estimate using $m_0\approx 0.9$ GeV gives
\begin{equation}
\label{eq:naive}
\delta \approx 0.2
\end{equation}

\section{QUENCHED LOGS IN THE PION MASS}

It has recently been argued \cite{Mawhinney} that the behavior of the pion 
mass as a function of the bare quark mass calculated in quenched lattice QCD 
shows little or no evidence for the presence of chiral logs at the level suggested
by the estimate (\ref{eq:naive}). We have analyzed 
the quenched pion mass from ACPMAPS data 
at four different $\beta$ values and a variety of hopping
parameters, as shown in Table 1. The pion masses were extracted from the
pseudoscalar propagator with pointlike sources. To determine and remove the effect of excited
states, both one- and two-exponential fits for a variety of  time windows were carried
out. For some of the $\kappa$ values at $\beta=5.7$,
the pion mass obtained in this way was compared with that obtained using
smeared-source quark propagators, and the results were found to agree within
statistical errors. For each $\beta$ value, the pion masses for $N$ values of $\kappa$
were calculated (here $N=3$ or $4$), with the full $N\times N$ error matrix being
computed by a jackknife elimination. By minimizing the covariant $\chi^2$,
a 3-parameter fit was obtained to the fitting function
\begin{equation}
\label{eq:fit}
m_{\pi}^2 = C \left(\kappa^{-1} - \kappa_c^{-1}\right)^{\frac{1}{1+\delta}}
\end{equation}
with fit parameters $C, \kappa_c,$ and $\delta$.
The results for the parameter $\delta$ are given in Table 1. For all four
values of $\beta$, the value obtained for $\delta$ is consistent with zero.
Combining the statistics of the four $\beta$'s, we get
\begin{equation}
\label{eq:limit}
\delta = 0.00 \pm .03
\end{equation}
In the analysis leading to (\ref{eq:limit}), only values of $\kappa$ corresponding to pion masses
of less than 750 MeV were included in the fits in an effort to minimize the 
effect of higher order chiral perturbation theory terms. In the $\beta=5.7$
result (where there were four mass points and hence one degree of freedom in the fit),
the covariant $\chi^2$ was 0.1, indicating that a good fit is obtained without the need
for higher order terms in (\ref{eq:fit}). 

\begin{table}
\caption{Exponent $\delta$ from $m_{\pi}^2$ vs. $m_q$.}
\label{tab:deltalimits}
\begin{tabular}{|c|c|c|}
\hline
$\beta$ & $\kappa$'s & $\delta$ \\
\hline
$5.7$ & $.161,.165,.1667,.168$ & $.015(47)$ \\
$5.9$ & $.157,.158,.159$ & $-.004(62)$ \\
$6.1$ & $.153,.154,.1545$ & $.009(75)$ \\
$6.3$ & $.1510,.1513,.1515$ & $-.033(114)$\\
\hline
\end{tabular}
\end{table}

\section{CALCULATION OF THE HAIRPIN DIAGRAM}

To investigate the apparent suppression of chiral logs further, we calculated the hairpin
diagram directly at $\beta = 5.7$, using the technique of Kuramashi,
et al \cite{Kuramashi}. The calculations were carried out on both $12^3\times 24$
and $16^3\times 32$ lattices. Our main conclusion is that {\it the value of $Am_0^2$
extracted from a direct calculation of the hairpin graph is much smaller than the
physical $\eta'$ mass-squared, and consistent with the small value of $\delta$
inferred from the limits on quenched chiral logs in the pion mass}. At the $\kappa$
values and lattice sizes for which a direct comparison could be made, our raw data
was in good agreement with that of Ref. \cite{Kuramashi}. Our somewhat different
conclusion regarding the suppression of the hairpin vertex at small $p^2$ follows from
several factors which we briefly mention here.
First, in order 
to extract the coefficient of the double pole 
term in the hairpin (i.e. $Am_0^2$), the pion propagators on either side of the vertex
were assumed to contain excited-state as well as ground-state contributions,
with the excited state mass determined from the previously described pion propagator analysis.
(The coefficient of the excited state term 
was a free parameter in the fit.) It is important to note that, in the hairpin
diagram, terms involving an excited state on one but not both sides of the hairpin
still contain a single Goldstone pole, and thus fall off with the same exponential factor
as the ground state term. They are only suppressed by a power of $t$, i.e. they 
fall off like $e^{-m_{\pi}t}$ instead of $te^{-m_{\pi}t}$. For our fits, the inclusion of excited state
contributions to the hairpin propagator reduced the extracted value of
$Am_0^2$ by about 30 to 50\% compared to a pure ground state fit.
Secondly, a comparison of our results for $12^3\times 24$ and $16^3\times 32$ lattices with
the same $\beta$ and $\kappa$ values exhibits a large and highly mass dependent
finite volume effect on the measured value of $Am_0^2$. This raises doubt about
the results of performing a chiral extrapolation on a fixed size box. (The chiral 
extrapolation contributed substantially to the quoted value of $m_0\approx 750$
MeV in Ref. \cite{Kuramashi}.) 
The results of the hairpin calculation for several $\kappa$'s and two box sizes
are converted to an effective value of $\delta$
and plotted in Fig. 1. They
are seen to be consistent with the one-standard-deviation upper bound from 
the pion mass analysis. (The dotted line represents the upper bound from the
$\beta = 5.7$ pion mass data in Table I.) 

\begin{figure}
\vspace*{4.6cm}
\includegraphics{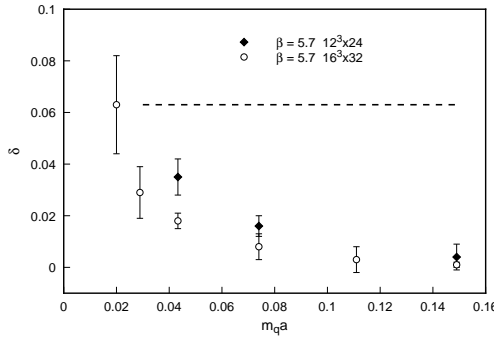}
\vspace{-0.5cm}
\caption[]{The value of the chiral log parameter $\delta$ extracted from the hairpin calculation
(using $\delta = Am_0^2/24\pi^2f_{\pi}^2)$ compared with the one standard deviation upper bound from
$m_{\pi}^2$ vs. $m_q$.}
\label{fig:effmass}
\end{figure}

For the range of quark masses considered, the results presented here provide strong evidence that the size of quenched chiral logs is suppressed by the fact that
the hairpin vertex evaluated on the pion mass shell is much smaller than expected from the assumptions that $m_0^2\approx m_{\eta'}^2$ and $A\approx 1$. If we discard 
the possibility that $m_0^2 << m_{\eta'}^2$ (which would be a disturbing failure of
QCD to reproduce the real world), it may be concluded that $A<<1$ and that the
hairpin is highly momentum dependent.
In addition to analyzing the time-dependence of the hairpin propagator, 
there is an indirect way to determine $m_0^2$ independently of $A$ by
appealing to the Witten-Veneziano formula, which relates $m_0^2$ to $\chi_t$, the 
topological susceptibility of pure glue,
\begin{equation}
\label{eq:WV}
m_0^2 = \frac{4N_f}{f_{\pi}^2}\chi_t
\end{equation}
It turns out that the same data generated in the hairpin calculation can also be used
to obtain an approximate measurement of the winding number of each gauge
configuration in an ensemble, using the anomalous chiral Ward identity,
\begin{equation}
\label{eq:Ward}
\int d^4x \langle \bar{\psi}\gamma^5 \psi \rangle_G = \frac{i\nu}{m_q}
\end{equation}
By studying the behavior of a single $\gamma^5$ loop integrated over the entire lattice as a
function of quark mass, (\ref{eq:Ward}) may be used to obtain an approximate
determination of the winding numbers of the configurations.
(A more detailed discussion of this method
and its applications will be presented elsewhere.)
From the same $\gamma^5$ loops used
to compute the hairpin at $\beta=5.7$ and $V = 12^3\times 24 a^4$, we obtain
a mean squared winding number of $\langle \nu^2 \rangle = 23 \pm 3$. The error
here is a very rough estimate based on varying the criteria for observing a
$1/m_q$ pole. From this, and using $a^{-1}=1.15$ GeV, we obtain the topological susceptibility
\begin{equation}
\chi_t = \langle \nu^2 \rangle/V \approx (180 MeV)^4
\end{equation}
and from the formula (\ref{eq:WV}), $m_0\approx 1.1\pm 0.2$ GeV. This is consistent with
results obtained by the cooling method\cite{Kuramashi,Hoek}.

We are grateful to George Hockney for many contributions to this research.

\end{document}